\documentclass[useAMS,usenatbib,twoside,a4paper]{mn2e}
\usepackage{graphicx}

\title[Binary period of SXP504]
{The binary period and outburst behaviour of the SMC X-ray binary
pulsar system SXP504.}
\author[W.R.T. Edge et al.]
       {W.R.T. Edge,$^1$ M.J. Coe,$^1$ J.L. Galache,$^1$ V.A. McBride,$^1$
       \newauthor R.H.D. Corbet,$^2{}^,{}^3$ A.T. Okazaki,$^4$ S.
Laycock,$^5$ C.B. Markwardt,$^2{}^,{}^6$ \newauthor F.E.
Marshall,$^2$ A. Udalski, $^7$
\\
        $^1$School of Physics and Astronomy, Southampton University, SO17
        1BJ, UK\\
        $^2$NASA Goddard Space Flight Center, Greenbelt, MD 20771 USA\\
        $^3$Universities Space Research Association\\
        $^4$Faculty of Engineering, Hokkai-Gakuen University, Sapporo, Japan\\
        $^5$Harvard-Smithsonian Center for Astrophysics, Cambridge, MA 02138, USA\\
        $^6$Department of Astronomy, University of Maryland, College Park, MD 20742,
        USA\\
        $^7$Warsaw University Observatory, Al. Ujazdowskie 4, 00-478 Warszawa,
        Poland}

\begin{document}

\date{Accepted 2005.
      Received 2005;
      in original form 2005}

\pagerange{\pageref{firstpage}--\pageref{lastpage}} \pubyear{2005}

\maketitle

\label{firstpage}

\begin{abstract}

A probable binary period has been detected in the optical
counterpart to the X-ray source CXOU J005455.6-724510 = RX
J0054.9-7245 = AXJ0054.8-7244 = SXP504 in the Small Magellanic
Cloud. This source was detected by Chandra on 04 Jul 2002 and
subsequently observed by XMM-Newton on 18 Dec 2003. The source is
coincident with an Optical Gravitational Lensing (OGLE) object in
the lightcurves of which several optical outburst peaks are visible
at $\sim268$ day intervals. Timing analysis shows a period of $268.6
\pm  0.1$ days at $>99{\%}$ significance. Archival Rossi X-ray
Timing Explorer (RXTE) data for the 504s pulse-period has revealed
detections which correspond closely with predicted or actual peaks
in the optical data. The relationship between this orbital period
and the pulse period of 504s is within the normal variance found in
the Corbet diagram.

\end{abstract}


\section{INTRODUCTION}

The Magellanic Clouds are a pair of satellite galaxies which are
gravitationally bound to our own but which have structural and
chemical characteristics differing significantly from each other,
and from the Milky Way. These differences are likely to be reflected
in the properties of different stellar populations. The Small
Magellanic cloud (SMC) is located at a distance of about 60 kpc
(Harries, Hilditch, \& Howarth 2003) and centred on a position of
R.A. 1hr Dec. -73$^{o}$. It is therefore close enough to be observed
with modest ground based telescopes whilst at the same time
providing an opportunity to study and compare the evolution of other
galaxies.

Intensive X-ray satellite observations have revealed that the SMC
contains an unexpectedly large number of High Mass X-ray Binaries
(HMXB). At the time of writing, 47 known or probable sources of this
type have been identified in the SMC and they continue to be
discovered, although only a small fraction of these are active at
any one time because of their transient nature. All X-ray binaries
so far discovered in the SMC are HMXBs (Coe et al. 2005).

Most High Mass X-ray Binaries (HMXBs) belong to the Be class, in
which a neutron star orbits an OB star surrounded by a circumstellar
disk of variable size and density. The optical companion stars are
early-type O-B class stars of luminosity class III-V, typically of
10 to 20 solar masses that at some time have shown emission in the
Balmer series lines. The systems as a whole exhibit significant
excess flux at long (IR and radio) wavelengths, referred to as the
infrared excess. These characteristic signatures as well as strong
H$\alpha $ line emission are attributed to the presence of
circumstellar material in a disk-like configuration (Coe 2000,
Okazaki \& Negueruela 2001).

The mechanisms which give rise to the disk are not well understood,
although fast rotation is likely to be an important factor, and it
is possible that non-radial pulsation and magnetic loops may also
play a part. Short-term periodic variability is observed in the
earlier type Be stars. The disk is thought to consist of relatively
cool material, which interacts periodically with a compact object in
an eccentric orbit, leading to regular X-ray outbursts. It is also
possible that the Be star undergoes a sudden ejection of matter
(Negueruela 1998, Porter \& Rivinius 2003).

Be/X-ray binaries can present differing states of X-ray activity
varying from persistent low or non-detectable luminosities to short
outbursts. Systems with wide orbits will tend to accrete from less
dense regions of the disk and hence show relatively small outbursts.
These are referred to as Type I (Stella et al., 1986) and usually
coincide with the periastron of the neutron star. Systems with
smaller orbits are more likely to accrete from dense regions over a
range of orbital phases and give rise to very high luminosity
outbursts, although these may be modulated by the presence of a
density wave in the disk. Prolonged major outbursts, which are not
confined to periastron passage, are normally called Type II
(Negueruela 1998).

\section{THE DATA}

\subsection{X-ray Data}
\label{sect:xray}

This X-ray object was discovered in archive data using an
observation made by the Chandra X-ray Observatory on 4 July 2002
(MJD 52459) (Edge et al. 2004). Its position was determined as R.A.
00:54:55.8 Dec. -72:45:11 to an accuracy of $\sim$1 arcsec . It was
found to have a period of $503.5 \pm 6.7 s$ at a $>99$\% level of
confidence. The source was subsequently independently identified by
Haberl et al. (2004) in an XMM-Newton observation of 18 Dec 2003
(MJD 52991). They computed the pulse period at $499.2 \pm 0.7s$. The
object is very close to RX J0054.9-7245 = AX J0054.8-7244 which is
listed by both Haberl \& Pietsch (Haberl \& Pietsch 2004) and
Yokogawa et al. (Yokogawa et al., 2003) as a HMXB pulsar candidate.

\begin{figure}
\begin{center}
\includegraphics[width=85mm]{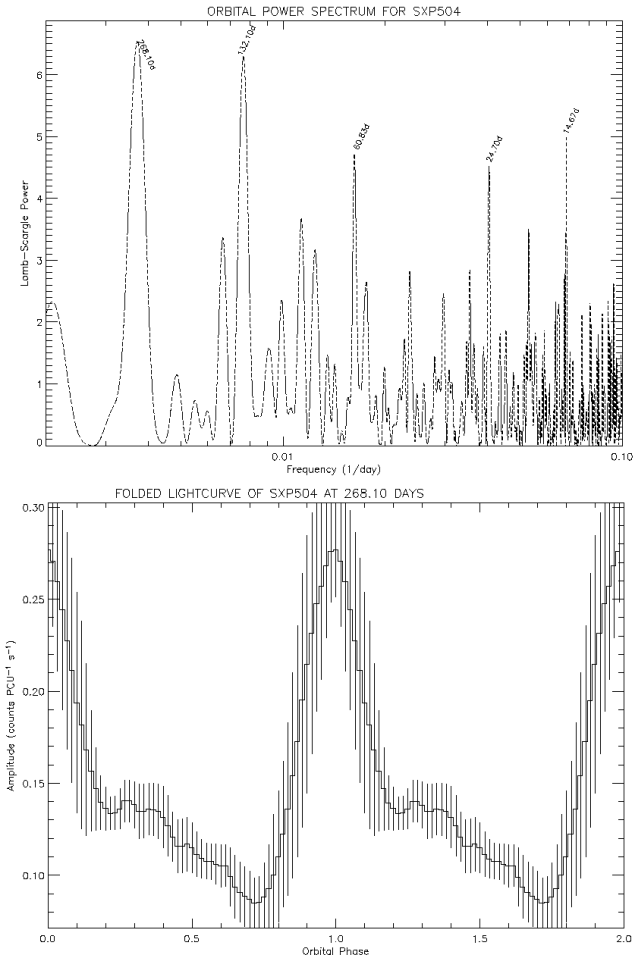}
\end{center}
\caption{Upper panel. Power spectrum from timing analysis of
consolidated RXTE observations of SXP 504.This shows a strong peak
at $268.1\pm5$ days. Lower panel. Lightcurve folded at 268.1 days.
$T_{0}$ is MJD 50560.} \label{fig:Xray}
\end{figure}

The Rossi X-ray Timing Explorer (RXTE) has been regularly monitoring
the SMC since 1997 on a weekly basis (Corbet et al. 2004, Laycock et
al. 2004). A search of archival RXTE Proportional Counter Array
(PCA) data for the 504s pulse-period revealed a considerable number
of detections. Those of $>99\%$ significance are listed in Table
~\ref{tab1} and are also plotted as histograms in Figures
~\ref{fig:ogle} and ~\ref{fig:macho}. Timing analysis was carried
out on the consolidated RXTE observations and revealed a strong peak
in the power spectrum at $268.1\pm5$ days at a significance of
56.5\%. This is shown in the upper panel of figure ~\ref{fig:Xray}.
The folded pulse profile of the RXTE X-ray lightcurve, is in the
lower panel. $T_{0}$ was put at MJD 50560.

\begin{table}
\begin{tabular}
{|p{48pt}|p{71pt}|p{20pt}|} \hline MJD& Flux&
Sig \\
~& (cts pcu$^{ - 1}$ s$^{ - 1})$&
({\%}) \\
\hline
51439.33& 0.330&
99.5 \\

51624.15& 0.741&
99.9 \\

51646.91& 1.108&
99.9 \\

51898.86& 0.899&
99.9 \\

51933.35& 0.544&
99.9 \\

51941.38& 0.495&
99.5 \\

52066.8& 0.700&
99.4 \\

52171.31& 0.639&
99.7 \\

52193.62& 0.964&
99.9 \\

52438.72& 1.336&
99.9 \\

52483.84& 0.732&
99.5 \\

52515.41& 0.915&
99.9 \\

52555.42& 0.724&
99.9 \\

52606.21& 0.583&
99.9 \\

52612.3& 0.686&
99.9 \\

52660.25& 0.930&
99.9 \\

52676.25& 0.495&
99.8 \\

52683.09& 0.715&
99.9 \\

52687.9& 0.749&
99.9 \\

52717.96& 0.720&
99.9 \\

52746& 0.551&
99.9 \\

52778.87& 0.678&
99.7 \\

52836.83& 0.895&
99.9 \\

52954.72& 0.474&
99.8 \\

52969.16& 0.851&
99.9 \\

53044.37& 0.451& 99.9 \\
\hline
\end{tabular}
\caption{RXTE detections of $>99\%$ significance, showing MJD, flux
and significance.} \label{tab1}
\end{table}

This object is likely to be the same as the ROSAT source RX
J0054.9-7245 which was detected in three observations: on 9-12 May
1993, on 15 Apr 1994 and 3 Apr - 2 May 1997. It was not detected in
several other ROSAT observations, however the detection threshold of
these was not as low as the May 1993 level. It may also be the ASCA
source AXJ0054.8-7244 which was detected in Nov 1998 (Haberl et al.
2004).

\subsection{Optical Data}

The position of the source given in section ~\ref{sect:xray}
coincides with the emission line object [MA93] 809 (Meyssonnier \&
Azzopardi, 1993) which is taken to be the optical counterpart. The
star has a V magnitude of 14.99 and a B-V colour index of -0.02 (Coe
et al. 2005) and appears in both the OGLE and MACHO databases. These
databases provide an opportunity to investigate the variability of
this object over a period of about 11 years.

\subsubsection{OGLE}

The Optical Gravitational Lensing Experiment (OGLE) is a long term
project, started in 1992, with the main goal of searching for dark
matter with microlensing phenomena (Udalski et al., 1998). Two sets
of OGLE data, designated II and III, are available for this object.
Both show I-band magnitudes using the standard system, however the
more recent OGLE III data have not yet been fully calibrated to
photometric accuracy. The source is coincident with the OGLE object
numbered 47103, in Phase II, and 36877, in Phase III.

\begin{figure}
\includegraphics[width=90mm]{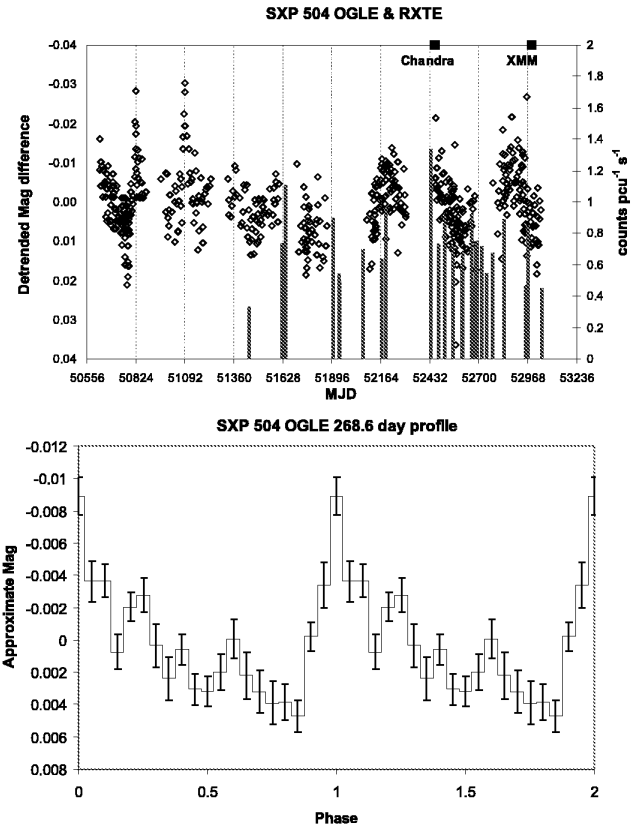}
\caption{SXP 504 OGLE lightcurve (Top panel). The epochs of the
Chandra and XMM-Newton observations are marked on the upper X axis
of the diagram. All RXTE detections of $>99\%$ significance are
shown against the right hand Y axis scale. Optical outburst peaks
are visible at MJD 50829, 51098 and 52965. These observations
indicate an orbital period of $\sim268$ days, marked on the X axis.
The bottom panel shows the OGLE lightcurve folded at 268 days using
the MJD 50556 zeropoint. The modulation amplitude is 0.015 mag.}
\label{fig:ogle}
\end{figure}

These lightcurves are shown in the top panel of figure
~\ref{fig:ogle}. An inspection of the raw data showed that the
partially calibrated Phase III data was offset from the Phase II
data by 0.05 mag. An adjustment of this amount was therefore applied
after which the lightcurves were joined and detrended with a 6th
order polynomial. The epochs of the Chandra and XMM-Newton
observations are marked on the upper X axis of the diagram. All RXTE
detections of $>99\%$ significance are also shown, the height of the
columns indicates the flux in $counts~pcu^{-1}~s^{-1}$ against the
right hand Y axis scale. Several optical outburst peaks are visible
at $\sim $268 day intervals, which are shown on the X axis of the
diagram.

In order to examine the profile, the lightcurve was folded at 268
days using the MJD 50556 zeropoint. The result is shown in the
bottom panel of figure ~\ref{fig:ogle} which reveals a sharp profile
with a peak to peak modulation of $0.015\pm0.002$ mag.

\subsubsection{MACHO}

In 1992 the  MAssive Compact Halo Objects project (MACHO) began a
survey of regular photometric measurements of several million
Magellanic Cloud and Galactic bulge stars (Alcock et al. 1993). The
MACHO data cover the period July 1992 to January 2000 and consist of
lightcurves in two colour bands described as \textit{blue} and
\textit{red}. \textit{Blue} is close to the standard $V$ passband
and \textit{Red }occupies a position in the spectrum about halfway
between $R$ and $I$ (Alcock et al. 1999).

\begin{figure}
\begin{center}
\includegraphics[width=90mm]{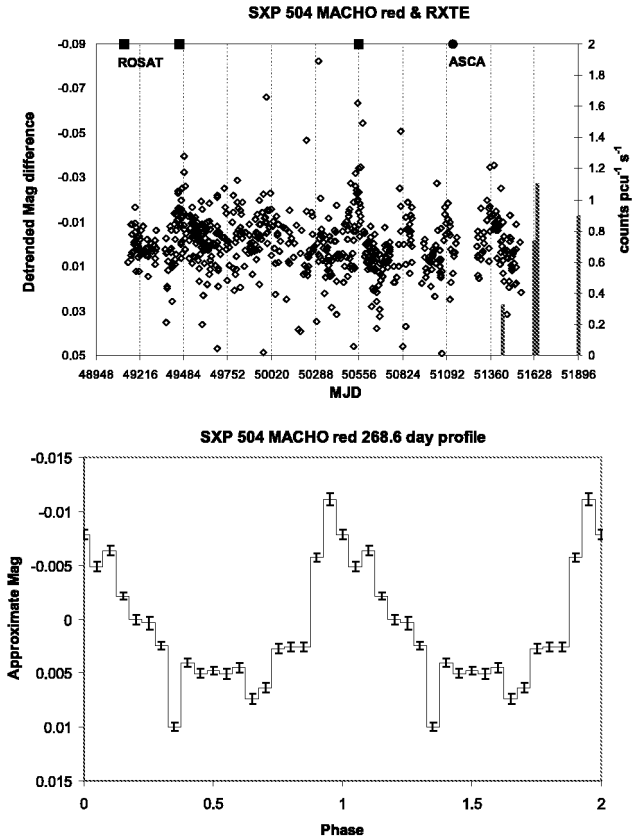}
\end{center}
\caption{SXP 504 MACHO \textit{red} lightcurve (Top panel). RXTE
detections of $>99\%$ significance are shown against the right hand
Y axis scale. The epochs of the ROSAT detections are shown as solid
squares and that of the ASCA detection as a solid circle, on the
upper X axis. there is clear evidence for optical outbursts at the
$\sim268$ day intervals marked on the X axis. The bottom panel shows
the MACHO \textit{red} lightcurve folded at 268 days using the MJD
50556 zeropoint. The modulation amplitude is 0.018 mag.}
\label{fig:macho}
\end{figure}

This source is coincident with MACHO object 207.16245.16. The
lightcurve from the \textit{red} data is shown in the top panel of
figure ~\ref{fig:macho}. A single rogue observation at MJD 50359,
which was 2 magnitudes brighter that all the others, was removed
after which it was detrended using a 5th order polynomial. The
figure shows clear evidence for optical outbursts at the $\sim268$
day intervals marked on the X axis. RXTE detections of $>99\%$
significance are shown against the right hand Y axis scale. The
epochs of the ROSAT (solid squares) and ASCA (solid circle)
detections are also shown on the upper X axis.

To examine the pulse profile, the lightcurve was also folded at 268
days (figure ~\ref{fig:macho}, bottom panel) using the same
zeropoint as  figure ~\ref{fig:ogle} (MJD 50556). The amplitude of
the peak to peak modulation is $0.018\pm0.0006$ mag.

The MACHO \textit{blue} lightcurve is not strongly modulated and did
not produce either a significant period or a coherent pulse profile.

\subsubsection{Optical timing analysis}

\begin{figure}
\begin{center}
\includegraphics[width=90mm]{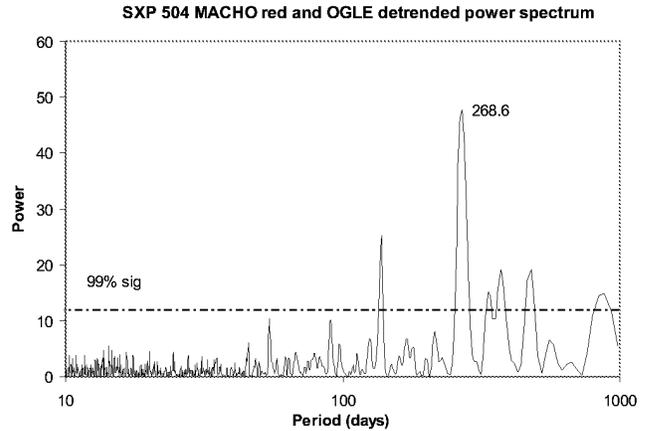}
\end{center}
\caption{SXP 504 combined MACHO \textit{red} and OGLE optical power
spectrum generated using the Lomb-Scargle algorithm. There is a
strong period at $268.6 \pm 0.1$ days which corresponds closely with
the observed outburst intervals. The smaller peak is the half-period
harmonic.} \label{fig:combpow}
\end{figure}

For the purpose of timing analysis The OGLE and MACHO \textit{red}
data were normalised and combined into a single dataset. This was
then subjected to Lomb-Scargle analysis which revealed a strong
period of $268.6 \pm 0.1$ days corresponding closely with the
observed outburst intervals. The power spectrum is in figure
~\ref{fig:combpow}. The smaller peak is the half-period harmonic.

\section{Discussion and Conclusion}

SXP504 has been identified as an X-ray binary pulsar system in the
SMC (Edge et al. 2004). It appears to be typical of such systems
insofar as the optical counterpart is a Be star, but it is unusual
because it is one of a minority of such objects that show visible
peaks in the optical lightcurves at the binary period. The most
comparable system in the SMC is SXP756 which also has a long orbital
period at 394 days as well as highly visible, but narrow and short
lived, optical outbursts.

The optical modulation is thought to arise when the neutron star in
a highly eccentric orbit briefly interacts with Be star disk at
periastron. The tidal torque, which is strongest at this point,
removes angular momentum from the outermost part of the disk,
causing it to shrink and its density to increase. In addition, the
two-armed spiral wave, which is excited at periastron, also enhances
the disk density. If the disk is optically thin, the luminosity will
also increase because the local emissivity is proportional to the
square of the density. This occurs just after the periastron passage
of the neutron star (Okazaki et al. 2002). The higher the orbital
eccentricity, the more rapid and significant the luminosity increase
is expected to be (Okazaki 2005, in preparation). The subsequent
decay of the optical outburst results from expansion of the disk by
viscous diffusion and is necessarily slower.

It follows that where the binary orbit is long and highly eccentric,
the disk is almost unaffected for much of the orbital phase and the
optical outbursts are likely to be more clearly marked.

The 268 day period detected in this system is visible as outburst
peaks in both the OGLE and MACHO lightcurves. The existence of these
peaks has been confirmed by folding the data and revealing the pulse
profiles. Lomb-Scargle analysis of the MACHO and OGLE data has
detected a period of $268.6 \pm 0.1$ days, which agrees closely with
the observed period of the outburst intervals.

All these observations can be described by an ephemeris of:

\begin{equation}\label{eqn:eph}
$T= (MJD 50556 $\pm $ 3) + n (268 $\pm $ 0.6)$
\end{equation}

where T is the epoch of the outburst and n is the outburst cycle
number.

Lomb-Scargle analysis of the consolidated RXTE observations has
detected a period of $268.1\pm5$ days. Furthermore the stronger
X-ray detections of $>99\%$ are nearly all very close to the peaks
predicted by the optical ephemeris, but because of non-continuous
X-ray coverage, others may well have been missed. The relationship
between this orbital period and the pulse period of 504s is within
the normal variance found in the Corbet diagram (Corbet 1984).

These results, taken together, confirm that 268 days is the binary
period of the system. They also provide an instructive example of
the use of parallel and complementary long term X-ray and optical
data in determining the orbital characteristics of an X-ray binary
system.

\section*{Acknowledgments}

This paper utilizes public domain data obtained by the MACHO
Project, jointly funded by the US Department of Energy through the
University of California, Lawrence Livermore National Laboratory
under contract No. W-7405-Eng-48, by the National Science Foundation
through the Center for Particle Astrophysics of the University of
California under cooperative agreement AST-8809616, and by the Mount
Stromlo and Siding Spring Observatory, part of the Australian
National University.

Partial support to the OGLE project was provided with the following
grants: Polish MNII grant 2P03D02124,  NSF grant AST-0204908 and
NASA grant NAG5-12212.

\end{document}